\documentclass[10pt]{article}

\begin{document}

\title{\Large \bf Supersymmetry in the Dirac equation for
generalized Coulomb potential
\author{\large Tamar~T.~Khachidze$^*$, Anzor~A.~Khelashvili$^*$ \bigskip \\
{\it  $^*$~ Department of Theoretical Physics, Ivane Javakhishvili
 Tbilisi State University,}
\\ {\it I.Chavchavadze ave. 3, 0128, Tbilisi, Georgia} \\ } }

\maketitle

{\large

{\bf Abstract}\\
\medskip
 We propose a symmetry of the Dirac equation
under the interchange of signs of eigenvalues of the Dirac's $K$
operator. We show that the only potential which obeys this
requirement is the Coulomb one for both vector and scalar cases.
Spectrum of the Dirac Equation is obtained algebraically for
arbitrary combination of
                  Lorentz-scalar and Lorentz-vector Coulomb potentials using the Witten's Superalgebra
                  approach. The results coincides with that, known from the explicit solution of the Dirac
                  equation.


\section{Introduction} \label{s1}

 The Kepler problem has an additional symmetry,
more precisely, additional conserved quantity, so-called
Laplace-Runge-Lenz (LRL) vector (for relevant references see, for
example,~ \cite{O'Connel2003}). Because of this fact there appears
extended algebra together with orbital momentum, which is
isomorphic to $SO(4)$ (for negative total energies). But
components of LRL vector as generators of this algebra do not
participate in any geometric transformations. In 1926
W.Pauli~\cite{Pauli1926} considered this algebra in quantum
mechanics and obtained Hydrogen atom spectrum by only algebraic
methods. In 30-ies V.Fock~\cite{Fock1935} considered the
Schroedinger equation for the Kepler problem in momentum
representation and had shown that the Coulomb spectrum has an
$SO(4)$ symmetry in energy-momentum space. Afterwards algebraic
methods attract more wide applications and it was cleared up in
the last decades that the hidden symmetry of the Kepler problem is
closely related to the supersymmetry of Hydrogen
atom~\cite{Stahlhofen1997,Dahl1995}

 The aim of our paper is to determine what the Dirac equation tells
 us about this problem. We can see that very interesting physical
 picture emerges.

   For the Kepler problem in the Dirac equation Johnson and
   Lippmann ~\cite{Johnson1950} published very brief abstract in
   Physical Review at 1950. They had written that there is an
   additional conserved quantity
\begin{equation}
A=\vec{\sigma}\cdot \vec{r}r^{-1} - \imath (\frac{\hbar
c}{e^2})(mc^2)^{-1}j\rho_{1}(H-mc^2\rho_3),
\end{equation}
 which plays the same role in Dirac equation, as the LRL vector
  in Schroedinger equation.

   As regards to the more detailed derivation,
 to the best our knowledge,  is not published in scientific literature
 (one of the curious fact in the history of physics of 20th
century)
      Moreover as far as commutativity    of the Johnson-Lippmann(JL) operator
       with the Dirac Hamiltonian is concerned, it is usually mentioned
      that it can be proved by  "rather tedious  calculations."~\cite{Katsura2006}.

     In papers~\cite{Khachidze2005,Khachidze2006}, we developed rather
     simple and transparent way for deriving the JL operator. We
     obtained this operator and at the same time proved its
     commutativity with the Hamiltonian.

      After we consider necessity to be convinced, that the Coulomb problem
      is distinguishable in this point of view. We considered the
      Dirac equation in arbitrary central potential, $V(r)$ and
      shown that the symmetry, which will be defined more precisely
      below, takes place only for Coulomb potential.

       Moreover it is remarkable that this property remains valid
       for arbitrary linear combination of Lorentz-vector and
       Lorentz-scalar potentials~\cite {Leviatan2004}. Therefore we expect that the
       spectrum for this general case can be obtained by pure
       algebraic methods, i.e. without solving of corresponding
       Dirac equation.

 Our paper is organized in the following manner: First of all we
 define a symmetry, which in fact leads to Witten's superalgebra.
 Then we determine the general class of so-called $K$-odd operators
 and apply this result to pure vector and then to mixed cases. At
 the end we derive the spectrum by algebraic methods on the ground
 of supersymmetric approach. We show that the final expression for
 ground state energy of Hydrogen atom coincides with that obtained
 by solving of equations of motion.

 \section{Survey Of Symmetry}

            Let consider the Dirac Hamiltonian in arbitrary central
            field, $\hat{V}(r)$, which in general may be Lorentz-scalar or 4th component of the
            Lorentz vector (or their combination)

\begin{equation}
H = \vec{\alpha}\cdot \vec{p}+\beta m +\hat{V}(r)
\end{equation}

            The so-called Dirac's $K$-operator, defined as
            \begin{equation}
K = \beta (\vec{\Sigma} \cdot \vec{l} +1)
\end{equation}
commutes with this Hamiltonian for arbitrary $\hat{V}(r)$,
$[K,H]=0$. Here $\vec{\Sigma}$ is the spin matrix
 $$
 \vec {\Sigma}  = \left( \begin{array}{l}
 \vec {\sigma} \ 0 \\
 0 \ \vec {\sigma}  \\
 \end{array} \right) \
$$
            and    $\vec{l}$    - orbital momentum vector-operator.

               It is evident that the eigenvalues of Hamiltonian
               also are labelled by eigenvalues of $K$. For
               example, the well-known \textbf{Sommerfeld formula} for the
               Coulombic spectrum looks like~\cite{Sommerfeld1951}
$$ E_{n,|k|}= m \left[
1+\frac{(Z\alpha)^2}{(n-|\kappa|+\sqrt{\kappa^2-(Z \alpha)^2})^2}
\right]^{-1/2} $$

               It manifests explicit dependence on $\kappa$  , more
               precisely on $|\kappa|=j+1/2$.
                  It is surprising that for other solvable
                  potentials the degeneracy with respect to signs
                  of     $\kappa$   does not take place.
   Physically this degeneracy leads to the forbidden of the Lamb
   shift. Indeed, positive $\kappa=j+1/2$ corresponds to aligned spin
   $j=l+1/2$, i.e. to states $(s_{1/2},p_{3/2}, \emph{etc})$, while
   negative $\kappa=-(j+1/2)$ corresponds to unaligned spin $j=l-1/2$,
   i .e. to states $(p_{1/2},d_{3/2},\emph{etc})$. So the absence of the
   Lamb shift $(s_{1/2}-p_{1/2})$ is a consequence of above mentioned
   degeneracy $\kappa \rightarrow -\kappa$.
   Naturally for description of this degeneracy one has to find an
   operator that mixes these two signs.
     It is clear that such an operator, say $A$,must be
     anticommuting with $K$:

\begin{equation}
\{A, K\} = AK+KA = 0
\end{equation}

              If at the same time this operator should commute
              with the Hamiltonian, then it'll generate the
              symmetry of the Dirac equation.
                Therefore, we need an operator $A$ with the
                following properties:
\begin{equation}
\{A, K\} = 0  ~~ {and} ~~[A, H]=0
\end{equation}

                  \textbf{It is our definition of symmetry we are looked
                   for.}

             It is worth to mention that after this operator is
             constructed, we will be able to define relativistic
             supercharges as follows:
$$Q_1=A, ~~~ Q_2 = \imath \frac{AK}{|\kappa|}$$

It is obvious that

$$\{Q_1,Q_2\} = 0, ~~~~ Q_1^2=Q_2^2 $$
 and we can find Witten's superalgebra~\cite{Witten1981}, where $Q_1^2 \equiv h$ is
a so-called, Witten's Hamiltonian.

 \section{Symmetry Operators}

  Now our goal is a construction of the operator $A$. We know, that there is a Dirac's $\gamma^5$ matrix,
   that anticommutes with $K$. \textbf{What else?} There is a \textbf{simple theorem}
    ~\cite{Khachidze2006}, according to which
    arbitrary $(\vec{\Sigma} \cdot \vec{V})$
\textbf{type operator}, \textbf{where} $\vec{V}$    \textbf{is a
vector with respect of} $\vec{l}$ and \textbf{is perpendicular to
it}, $(\vec{l}\cdot \vec{V})=(\vec{V} \cdot \vec{l})$,
\textbf{anticommutes with} $K$:

\begin{equation}
\{(\vec{\Sigma}\cdot \vec{V}), K\} = 0
\end{equation}
It is evident that the class of operators anticommuting with $K$
(so-called $K$-odd operators) is much wider.
 Any operator of the form    $\hat{O}(\vec{\Sigma}\cdot \vec{V})$ ,
 \textbf{where $\hat{O}$
commutes with $K$, but is otherwise arbitrary, also is a $K$-odd.}
This fact will be used below.

 \subsection{Pure Vector Potential}

   Let first establish the form of symmetry operator for vector
   potential only $\hat{V}(r)=V(r)$. By using the above mentioned
   theorem we wish to construct the $K$-odd operator $A$, that
  commutes with $H$. It is clear that there remains large freedom according to the above mensioned remark about
   $\hat{O}$ operators - one can take $\hat{O}$ into account or ignore it.

   We have the following physically interesting vectors at hand which obey the requirements of our theorem. They are
     \begin{equation}   \hat{\vec{r}}  - \textbf{unit radius-vector  and}   ~~   \vec{p}     - \textbf{linear momentum
     vector} \end{equation}
Both of them are perpendicular to $\vec{l}$.  Constraints of this
theorem are also satisfied by LRL vector
$\vec{A}=\hat{\vec{r}}-\frac\imath{2ma}[\vec{p}\times\vec{l}-\vec{l}\times\vec{p}]$,
but this vector is associate to the Coulomb potential. Hence we
abstain from its consideration for now.
 Thus, we choose the following $K$-odd terms
\begin{equation} (\vec{\Sigma}\cdot \hat{\vec{r}}), ~~~
K(\vec{\Sigma}\cdot \vec{p}) ~~~ {and} ~~~ K \gamma^5
\end{equation}
and let probe the combination
\begin{equation} A=x_1(\vec{\Sigma}\cdot \hat{\vec{r}})+\imath x_2
K(\vec{\Sigma}\cdot \vec{p}) + \imath x_3 K \gamma^5 f(r)
\end{equation}

Here the coefficients   $x_i(i=1,2,3)$ are chosen in such a way
that $A$ operator is Hermitian for arbitrary real numbers and
$f(r)$ is an arbitrary scalar function to be determined later from
the symmetry requirements. Let's calculate
\begin{eqnarray}
0=[A,H]=(\vec{\Sigma}\cdot \hat{\vec{r}})\{x_2 V'(r)-x_3 f'(r)\}
+\\ +\nonumber{2 \imath \beta K \gamma^5 \{\frac{x_1}{r} - m x_3
f(r)\}}
\end{eqnarray}

  We have a diagonal matrix in the first term, while the antidiagonal matrix in the second one.
  Therefore two equations follow:
  \begin{equation}
  x_2 V'(r) = x_3 f'(r),~~~~
  x_3 m f(r)=\frac{x_1}{r}
  \end{equation}

One can find from these equations:
\begin{equation}
V(r)= \frac{x_1}{x_2} \frac{1}{mr}
\end{equation}
\textbf{Therefore only the Coulomb potential corresponds to the
above required $\pm \kappa$ degeneracy.}
   The final form of obtained $A$ operator is the following:
\begin{equation}
A=x_1 \{(\vec{\Sigma}\cdot \hat{\vec{r}})-
\frac{\imath}{ma}K(\vec{\Sigma}\cdot \vec{p}) + \frac{\imath}{mr}K
\gamma^5\}
\end{equation}
where unessential common factor $x_1$ may be omitted and after using
known relations for Dirac matrices, this expression may be reduced
to the form
\begin{equation}
A = \gamma^5 \{ \vec{\alpha} \cdot \vec{r} - \frac{\imath}{m a}K
\gamma^5(H- \beta m)\}
\end{equation}

Above and here $a$ is a strengh of Coulomb potential, $a=Z
\alpha$. Precisely this operator is given in Johnson's and
Lippmann's abstract~\cite{Johnson1950}.

\textbf{What the real physical picture is standing behind this?}
Taking into account the relation

\begin{equation}
K \left(\vec{\Sigma}\cdot\vec{p}\right)=
 - \imath \beta \left( \vec{\Sigma} \cdot
\frac{1}{2}[\vec{p} \times \vec{l} - \vec{l} \times
\vec{p}]\right)
\end{equation}
one can recast our operator in the following form $$ A =
\vec{\Sigma}\cdot \left(\hat{\vec{r}} - \frac{\imath}{2ma} \beta
\left[\vec{p}\times\vec{l}-\vec{l}\times\vec{p}\right]\right)+\frac{\imath}{mr}K
\gamma^5 $$ One can see that in the non-relativistic limit, i.e.
$\beta\rightarrow 1$ and $\gamma^5 \rightarrow 0$ our operator
reduces to
\begin{equation}
A \rightarrow A_{NR}=
\vec{\sigma}\cdot\left(\hat{\vec{r}}-\frac{\imath}{2ma}[\vec{p}\times\vec{l}-
\vec{l}\times\vec{p}]\right)
\end{equation}

Note the LRL vector in the parenthesis of this equation.
\textbf{Therefore relativistic supercharge reduces to the
 projection of the LRL vector on the electron spin direction. Precisely this operator was used in the case of Pauli
electron~\cite {Tangerman1993}.}

  Because the Witten's Hamiltonian is
  \begin{equation}
  A^2=1+\left(\frac{K}{a}\right)^2 \left(\frac{H^2}{m^2}-1\right)
  \end{equation}
and it consists only mutually commuting operators, it is possible
their simultaneous diagonalization and replacement by
corresponding eigenvalues. For instance, the energy of ground
state is
\begin{equation}
E_0=\left(1-\frac{(Z\alpha)^2}{\kappa^2}\right)^{1/2}
\end{equation}
By using the ladder procedure, familiar for SUSY quantum
mechanics, the Sommerfeld formula can be easily
 derived~\cite{Katsura2006}.

 \subsection{Inclusion of Lorentz-scalar Potential}

  Let's remark that if we include the Lorentz-scalar potential as well
\begin{equation}
H=\vec{\alpha}\cdot \vec{p}+\beta m +V(r) + \beta S(r)
\end{equation}
this last Hamiltonian also commutes with $K$-operator, but
\textit{does not commute} with the above JL operator.

  On the other hand, non-relativistic quantum mechanics is \textbf{indifferent} with regard
  of \textbf{the Lorentz variance
   properties} of potential. Therefore it is expected that in case of scalar potential the description of
    hidden symmetry should also be possible. In other words, \textbf{the JL operator must be generalized} to this case.

   For this purpose it is necessary to increase number of $K$-odd structures. One has to use our theorem
   in the part of additional $\hat{O}$ factors.

  Now let us probe the following operator
  \begin{equation}
  X=x_1\left(\vec{\Sigma}\cdot\hat{\vec{r}}\right)+x_1'\left(\vec{\Sigma}\cdot\hat{\vec{r}}\right)H
  +\imath x_2 K \left(\vec{\Sigma}\cdot \vec{p} \right) + \imath
  x_3 K \gamma^5 f_1(r)+\imath x_3' K \gamma^5 \beta f_2(r)
  \end{equation}

We included factor $\hat{O}=H$ in the first structure and at the
same time  the matrix $\hat{O}=\beta$ in the third structure.
 Both of them commute with $K$ and are permissible by our theorem. The form (20) is a minimal extension of the
  previous picture, because only the first order structures in $\hat{\vec {r}}$ and $\vec{p}$ participate.
    For turning to the previous
   case one must take $x_1'=x_3'=0$ and $S(r)$=0.
Calculation of relevant commutators gives:
\begin{eqnarray}
[X,H] = \gamma^5 \beta K \{\frac{2\imath x_1}{r}- 2 \imath x_3 (m+S)
f_1(r)+\frac{2\imath x_1'}{r}V(r)\}+\\ \nonumber{ + K
(\vec{\Sigma}\cdot \hat{\vec{r}}) \{x_2 V'(r) - x_3 f_1'(r)\}}+\\
\nonumber{
 + K\beta \left( \vec{\Sigma}\cdot \hat{\vec{r}}\right) \{x_2 S'(r)-x_3' f_2'(r)\}+}
 \\
\nonumber{+ \gamma^5 K \{ \frac{2\imath x_1'(m+S)}{r}-2\imath x_3'
(m+S)f_2(r)\}+}\\ \nonumber{ + \beta K \{ \frac{2\imath x_1'}{r}-2
\imath x_3' f_2(r)\} (\vec{\Sigma} \cdot \vec{p})}
\end{eqnarray}
Equating this expression to zero,we derive matrix equation, then
passing to $2\times2$ representation we must equate to zero the
coefficients standing in fronts of diagonal and antidiagonal
elements. In this way it follows equations:

   \textbf{(i) From antidiagonal structures} ($\gamma^5,~ \gamma^5 \beta K$)

\begin{eqnarray}
~~\frac{x_1}{r}-x_3(m+S)f_1(r)+\frac{x_1'}{r}V(r)=0 \\
\nonumber{\frac{x_1'}{r}(m+S)-(m+S)f_2(r)=0}
\end{eqnarray}

   \textbf{(ii) From diagonal structures}( $K(\vec{\Sigma} \cdot \hat{\vec{r}}),~~K \beta
    (\vec{\Sigma}\cdot \hat{\vec{r}}), ~~ \beta K (\vec{\Sigma}\cdot \vec{p}) ~ $):
    \begin{eqnarray}
    x_2 V'(r)-x_3f_1'(r)=0\\
    \nonumber{\frac{x_1}{r}-x_2(m+S)V(r)+\frac{x_1'}{r}V(r)=0}\\
    \nonumber{x_2 S'(r)-x_3'f_2'(r)=0}\\
    \nonumber{\frac{x_1'}{r}-x_3' f_2(r)=0}
\end{eqnarray}

  Integrating the first and third equations in (23) for vanishing boundary conditions at infinity, we obtain
\begin{eqnarray}
f_1(r)=\frac{x_2}{x_3}V(r), ~~~~ f_2(r)=\frac{x_2}{x_3'}S(r)
\end{eqnarray}
and taking into account the last equation from (23), we have
\begin{equation}
f_2(r)=\frac{x_1'}{x_3' r}
\end{equation}
Therefore, according to (24), we obtain finally
\begin{equation}
S(r)=\frac{x_1'}{x_2 r}
\end{equation}

\textbf{So, the scalar potential must be Coulombic.}

 Inserting
Eq.(24) into the first equation of (22) and solving for $V(r)$,
one derives
\begin{equation}
V(r)=\frac{x_1}{r} \frac{1}{x_2(m+S)-\frac{x_1'}{r}}
\end{equation}
At last, using here the expression (26), we find
\begin{equation}
V(r) = \frac{x_1}{x_2 m r}
\end{equation}

 \textbf{Therefore we have asserted that the $\pm \kappa$ degeneracy is a symmetry of the Dirac equation only for Coulomb
 potential
 (for any general combination of Lorentz scalar and 4th component of a Lorentz
 vector).}

   It seems that the conservation of LRL vector is a macroscopic manifestation of symmetry, which is present
  in microworld.

  Further, if we take into account above obtained solutions, one can reduce the $X$ operator to more compact form
  \begin{equation}
  X = (\vec{\Sigma}\cdot\hat{\vec{r}})(ma_V+a_SH)-\imath K
  \gamma^5(H-\beta m)
  \end{equation}
where the following notations are used $$ a_V= - \frac{x_1}{x_2
m}, ~~~ ~~~ a_S= - \frac{x_1'}{x_2} $$
 Here $a_i$-s are the
constants of corresponding Coulomb potentials

$$V(r)=-\frac{a_V}{r},~~~~~~~S(r)=-\frac{a_S}{r} $$

 In conclusion we want to remark, that this expression for $X$ was obtained earlier by Leviatan~\cite{Leviatan2004}, who
 used the radial decomposition and separation of spherical angles in the Dirac equation.

   Our approach is 3-dimensional, without any referring to radial equation and, therefore is more systematic and rather easy.

 \section{Algebraic Derivation of Spectrum}

  Now we want to obtain spectrum of the
      Dirac Hamiltonian pure algebraically, without any referring
      on equations of motion. Our method is based on Witten's
      superalgebra,established in Sec.II above.

    Now we explore this algebra in a manner as in paper \cite{Katsura2006}.
     One defines a SUSY ground state $|0\rangle $:
          \begin{equation}
     h|0\rangle = X^2 |0\rangle =0~~\longrightarrow~~X|0\rangle =0
      \end{equation}
    Because  $X^2$ is a square of Hermitian operator, it has a positive definite spectrum and one
    is competent to take zero this operator itself in ground state. By this requirement we'll
     obtain Hamiltonian in this ground state and, correspondingly, ground state energy. After that by well
      known ladder procedure we can construct the energies of all excited levels. We believe that this method
      requires more strong justification, but nevertheless we are convinced, that it is true.

     Let equate $X=0$ and solve $H$ from Eq. (29):

      \begin{equation}
      H=m [(\vec{\alpha}\cdot \hat{\vec{r}})a_S + \imath
      K]^{-1} [\imath K \beta - a_V (\vec{\alpha}\cdot
      \hat{\vec{r}})]=\frac{m}{\kappa^2+a_S^2} N
      \end{equation}
where

 $$
    N \equiv [(\vec{\alpha}\cdot\hat{\vec{r}})a_S-\imath K]
    [\imath K\beta - a_V (\vec{\alpha}\cdot \hat{\vec{r}})]=
            $$

\begin{equation}
= -a_S a_V + K[ K\beta+\imath a_V
     ( \vec{\alpha} \cdot \hat{\vec{r}})] - \imath a_S K \beta
     (\vec{\alpha}\cdot\hat{\vec{r}})
     \end{equation}

 Now we try to diagonalize this operator using
Foldy-Wouthuysen ~\cite{Foldy1950} like transformation. Because the
second and third terms do not commute with each others we need
several (at least two) such transformations.

   We choose the first transformation in the following manner

    \begin{equation}
      \exp(\imath S_1)= \exp \left( -\frac{1}{2} \beta
      (\vec{\alpha}\cdot \hat{\vec{r}}) w_1 \right)
      \end{equation}

It is evident that

 \begin{equation}
\exp(\imath S_1) (\vec{\alpha}\cdot \hat{\vec{r}}) \exp(-\imath
S_1)=\exp(2\imath S_1)(\vec{\alpha}\cdot \hat{\vec{r}})
\end{equation}

$$ \exp(\imath S_1) \beta \exp(-\imath S_1) = \exp(2 \imath S_1)
\beta $$

Moreover

$$\exp(\imath S_1) K \exp(-\imath S_1)=K,$$
\begin{equation}
 \exp(\imath S_1) \beta
K \exp(-\imath S_1) = \exp(2 \imath S_1) \beta K
\end{equation}

and

    \begin{equation}
\exp(\imath S_1) \beta (\vec{\alpha}\cdot \hat{\vec{r}})
\exp(-\imath S_1)=\beta (\vec{\alpha}\cdot \hat{\vec{r}})
      \end{equation}

   Therefore the first transformation acts as

$$N' \equiv \exp(\imath S_1) N
 \exp(- \imath S_1) =$$
 \begin{equation}
 = - a_S a_V + K \exp(2\imath S_1 )
[K \beta + \imath a_V(\vec{\alpha} \cdot \hat{\vec{r}} )]
 -\imath a_S K\beta(\vec{\alpha}\cdot\hat{\vec{r}})
     \end{equation}

    But
     $\exp( 2\imath S_1)=
     \cosh {w_1}
     +\imath\beta(\vec{\alpha}\cdot\hat{\vec{r}})\sinh{w_1}$.
    Make use of this relation, we have

$$\exp(2\imath S_1) [K \beta +\imath a_V ( \vec{\alpha}\cdot
\hat{\vec{r}})]=$$
\begin{equation} \beta[K\cosh w_1 + a_V \sinh w_1]+ K
(\vec{\alpha}\cdot \hat {\vec{r}})[\imath a_V \cosh w_1 +\imath K
\sinh w_1 ]
\end{equation}
   Now in order to get rid of non-diagonal  $(\vec{\alpha}\cdot \hat {\vec{r}})$ terms, we must choose

\begin{equation}
\tanh w_1 = - \frac {a_V}{K}
\end{equation}

  Using simple trigonometric relations we arrive at
\begin{equation}
\exp(2\imath S_1)[K+\imath a_V (\vec{\alpha}\cdot \hat {\vec{r}})] =
K^{-1} \beta \sqrt {\kappa^2 - a_V^2}
\end{equation}

Let us perform the second F.-W. transformation

\begin{equation}
N''=\exp(\imath S_2) N' \exp(-\imath S_2),  ~~~~ where ~~~~ S_2 =
-\frac{1}{2}(\vec{\alpha}\cdot \hat {\vec{r}})w_2
\end{equation}

Now
 \begin{equation}
\exp(\imath S_2) K \beta \exp(-\imath S_2) = \exp(2\imath S_2) K
\beta
\end{equation}

$$\exp(\imath S_2) K \beta (\vec{\alpha}\cdot \hat
{\vec{r}})\exp(-\imath S_2) = \exp(2\imath S_2) K \beta
(\vec{\alpha}\cdot \hat {\vec{r}}) $$

$$ \exp(2\imath S_2) = \cos w_2 - \imath (\vec{\alpha}\cdot \hat
{\vec{r}}) \sin w_2$$
 Therefore

$$N'' = - a_S a_V +K \sqrt{\kappa^2 -
 a_V^2}\exp(2 \imath S_2 )\beta - $$
$$ - \imath a_S \exp(2\imath S_2) K \beta(\vec{\alpha}\cdot \hat
{\vec{r}})$$

$$ = -a_S a_V +K \sqrt{\kappa^2 - a_V^2}\beta \cos w_2 +\imath K
\sqrt{\kappa^2 - a_V^2} \beta(\vec{\alpha}\cdot \hat {\vec{r}}) \sin
w_2 -$$
\begin{equation}
 -\imath a_S K \beta (\vec{\alpha}\cdot \hat {\vec{r}})
\cos w_2 + a_S K \beta \sin w_2
\end{equation}

Requiring absence of  $(\vec{\alpha}\cdot \hat {\vec{r}})$ terms we
have

 \begin{equation}
\tan w_2 = \frac {a_S}{\sqrt{\kappa^2-a_V^2}}
\end{equation}

Therefore

 \begin{equation}
N'' = - a_S a_V + K \beta \sqrt{\kappa^2-a_V^2+a_S^2}
\end{equation}

and finally

 \begin{equation}
H = \frac{m}{\kappa^2+a_S^2}\{ -a_S a_V + K \beta
\sqrt{\kappa^2-a_V^2+a_S^2}\}
\end{equation}

For eigenvalues in ground state we have

 \begin{equation}
E_0 = \frac{m}{\kappa^2+a_S^2}\left[ -a_S a_V  \pm \kappa
\sqrt{\kappa^2-a_V^2+a_S^2}\right]
\end{equation}

Now let us remember the result obtained by explicit solution of the
Dirac equation for this case \cite{Greiner1985}

$$\frac{E}{m}= \frac{-a_S a_V}{a_V^2+(n-|\kappa|+\gamma)^2}$$
 \begin{equation}\pm
\sqrt{\left(\frac{a_S a_V}{a_V^2+(n-|\kappa|+\gamma)^2}\right)^2+
\frac{(n-|\kappa|+\gamma)^2-a_S^2}{a_V^2+(n-|\kappa|+\gamma)^2}},
\end{equation}

where
\begin{equation}
\gamma^2=\kappa^2-a_V^2+a_S^2
\end{equation}

     In the ground state  $n=1$, $j=1/2$ $\longrightarrow$ $|\kappa|=j+1/2=1$, there remains

     \begin{equation}
E_0=m\left[\frac{-a_S a_V}{a_V^2+\gamma^2} \pm \sqrt{\left(\frac{a_S
a_V}{a_V^2+\gamma^2}\right)^2 + \frac {\gamma^2 -
a_S^2}{a_V^2+\gamma^2}}\right],
\end{equation}
which after obvious manipulations reduces to our above derived
expression (47).
     Therefore by only algebraic methods we have obtained the correct expression for ground state energy.
     For obtaining of total spectrum it is sufficient now to use the Witten's algebra. Following the paper
      \cite{Katsura2006} ,
the ordinary ladder procedure consists in change (for our case):

$$ \gamma \longrightarrow \gamma + n - |\kappa| $$

    Making this, it follows from our lowest energy formula (47) the correct expression for total energy spectrum, Eq. (48).

    \section{Conclusions}

         In conclusion, by using of pure algebraic manipulations we obtained the spectrum of generalized Coulomb
    problem of the Dirac equation for arbitrary combination of Lorentz-scalar and Lorentz-vector potentials.
    This fact demonstrates a power of symmetry considerations. It is
    interesting by itself that the requirement of validity of
    Witten's algebra, or equivalently, ${S(2)}$ supersymmetry of the
    Dirac Hamiltonian for arbitrary central potentials (scalar or
    vector) leads to \emph{unique role of the Coulomb potential.\emph{}}

    \section*{Acknowledgments} The authors are indebted to express their
    gratitude to Prof. A.N.Tavkhelidze for many valuable discussions
    and critical remarks. We thank also Prof. Laszlo Jenkovszky for
    its permanent attention in the period and after the Yalta
    Conference, where this paper were reported.

     This work was supported by NATO Reintegration Grant No. FEL.
     REG. 980767.

}

\end{document}